\documentclass[a4paper,12pt]{article}

\usepackage[english]{babel}
\usepackage[utf8x]{inputenc}
\usepackage[T1]{fontenc}
\usepackage[]{natbib}
\usepackage{url}
\usepackage{listings}
\usepackage{mathtools}
\usepackage{chngcntr}
\usepackage{multirow}
\usepackage{authblk}
\usepackage{amsmath}
\usepackage{graphicx}
\usepackage{centernot}

\newcommand{\bigCI}{\mathrel{\text{\scalebox{1.07}{$\perp\mkern-10mu\perp$}}}}
\newcommand{\nbigCI}{\centernot{\bigCI}}

\usepackage[a4paper,top=3cm,bottom=2cm,left=3cm,right=3cm,marginparwidth=1.75cm]{geometry}

\usepackage{amsmath}
\usepackage{graphicx}
\usepackage[colorinlistoftodos]{todonotes}
\usepackage[colorlinks=true, allcolors=blue]{hyperref}

\title{Interpreting Hazard Ratios: Insights from Frailty Models.}

\author[1]{Mats Julius Stensrud\thanks{Corresponding author. Email: m.j.stensrud@medisin.uio.no}}
\affil[1]{\footnotesize Department of Biostatistics, Institute of Basic Medical Sciences, University of Oslo, Postbox 1122 Blindern, 0317 Oslo, Norway}

\begin{document}
\maketitle

\begin{abstract}
{Hazard ratios are often used to evaluate time to event outcomes, but they may be hard to interpret. A particular issue arise because hazards are typically estimated conditional on survival, i.e.\ on left truncated samples. Then, hazard ratios from conventional models cannot be interpreted as counterfactual hazard ratios that are immediately relevant to individual patients. This article explores how the hazard ratios from Cox models may differ from hazard ratios with a causal interpretation. Using summary data from twin studies, I suggest an approach to learn about the unmeasured heterogeneity in risk of an outcome, and this information allows us to explore the interpretation and magnitude of hazard ratios. Under explicit parametric assumptions, I present a two-step method to obtain hazard ratios that are more relevant to individual subjects. The strategy relies on untestable assumptions, but may nevertheless be useful for sensitivity analyses that are relatively easy to interpret.


}\end{abstract}
\clearpage
\counterwithout{equation}{section}
\counterwithout{table}{section}

\section{Introduction}
For time to event outcomes, hazard models are convenient because they allow for straightforward regressions on covariates. Nevertheless, it is well-known that estimates on the hazard scale, e.g.\ hazard ratios (HRs), are hard to interpret causally \citep{robins1989probability, greenland1996absence, hernan2010hazards, aalen2015does, stensrud2017exploring}. A particular issue arise due to left truncation: An exposure may be introduced at time $t_0$, but hazard estimates at a later time $t > t_0$ are calculated from subjects alive at $t$. Hence, at any $t > t_0$ the HRs are derived from left truncated samples, and not from the baseline population. Standard HR models, e.g.\ the Cox model, are based on multiplying likelihood functions for each event time, and thereby they are sequentially calculated from left truncated samples. Thus, such HRs do not have an immediate relevance for individual subjects \citep{robins1989probability}, even if the subjects are followed from the onset of an exposure. This issue has been denoted the inbuilt selection bias of the HR \citep{hernan2010hazards}, truncation bias \citep{vansteelandt2017survivor} or survival bias \citep{ aalen2015understanding}. 

The issue of left truncation is particularly severe if the onset of follow-up is delayed. In observational studies, exposures are often present before the subjects are recruited to the study, and the study sample is not necessarily representative of the pre-exposure population. Even in randomised controlled trials, the same issue arise when the treatment effect is assessed in individuals at a time $t$ later than randomisation at $t_0$ \citep{hernan2010hazards, greenland1996absence, aalen2015does, stensrud2017exploring,mcnamee2017serious}. In such scenarios, causal inference is not straightforward.

Intuitively, individuals who die before $t_0$ are expected to be the more \textit{frail}; the hazard of an event may be heterogeneously distributed across individuals, and the subjects who died before $t_0$ were expected to have higher average hazard \citep{vaupel1979impact, hougaard1995frailty, aalen2015understanding}. In some scenarios we have a good understanding of the factors that determine the individual risk, and then we may include these factors as covariates in our model. However, the heterogeneity may often be due to unobserved factors, and adjusting for measured factors will not be sufficient. In such situations, frailty models have been used to account for the variation in susceptibility to disease \citep{vaupel1979impact,lancaster1979econometric, moger2004frailty, haugen2009frailty, moger2008regression, moger2004analysis}. The frailty models may be seen as extensions of the Cox model, allowing for an unknown heterogeneity parameter. Unfortunately, specifying the parameters of the frailty distribution is not obvious, in particular when analysing data on independent individuals.

This article suggests an approach to learn about the frailty distribution using published summary data. After estimating the frailty distribution, it is possible to explore the causal interpretation of HRs. In particular, I describe a method to adjust for survival bias in proportional hazards models. The strategy consists of two steps: 1) a standard Cox proportional hazards model is fitted to obtain a marginal HR, and 2) this HR is adjusted to account for unmeasured heterogeneity, using frailty theory. The method relies on strong parametric assumptions, but under these conditions we may use published summary data from e.g.\ twin studies to find the frailty distribution, and thereby obtain frailty adjusted HRs, assuming that the frailty is shared among identical twins. 

I will illustrate that left truncation is an issue in genetic epidemiology, e.g.\ Mendelian Randomisation studies \citep{vansteelandt2017survivor,boef2015mendelian}. These studies rely on genetic variants that are carried from conception, but subjects are often included into these studies in late adulthood.  In particular, I will use published summary data on the relation between mortality and a gene associated with alcohol consumption to highlight the magnitude of bias.

\section{Issues with conditioning on survival.}
\begin{figure}
\centering
\includegraphics[width=4.5cm]{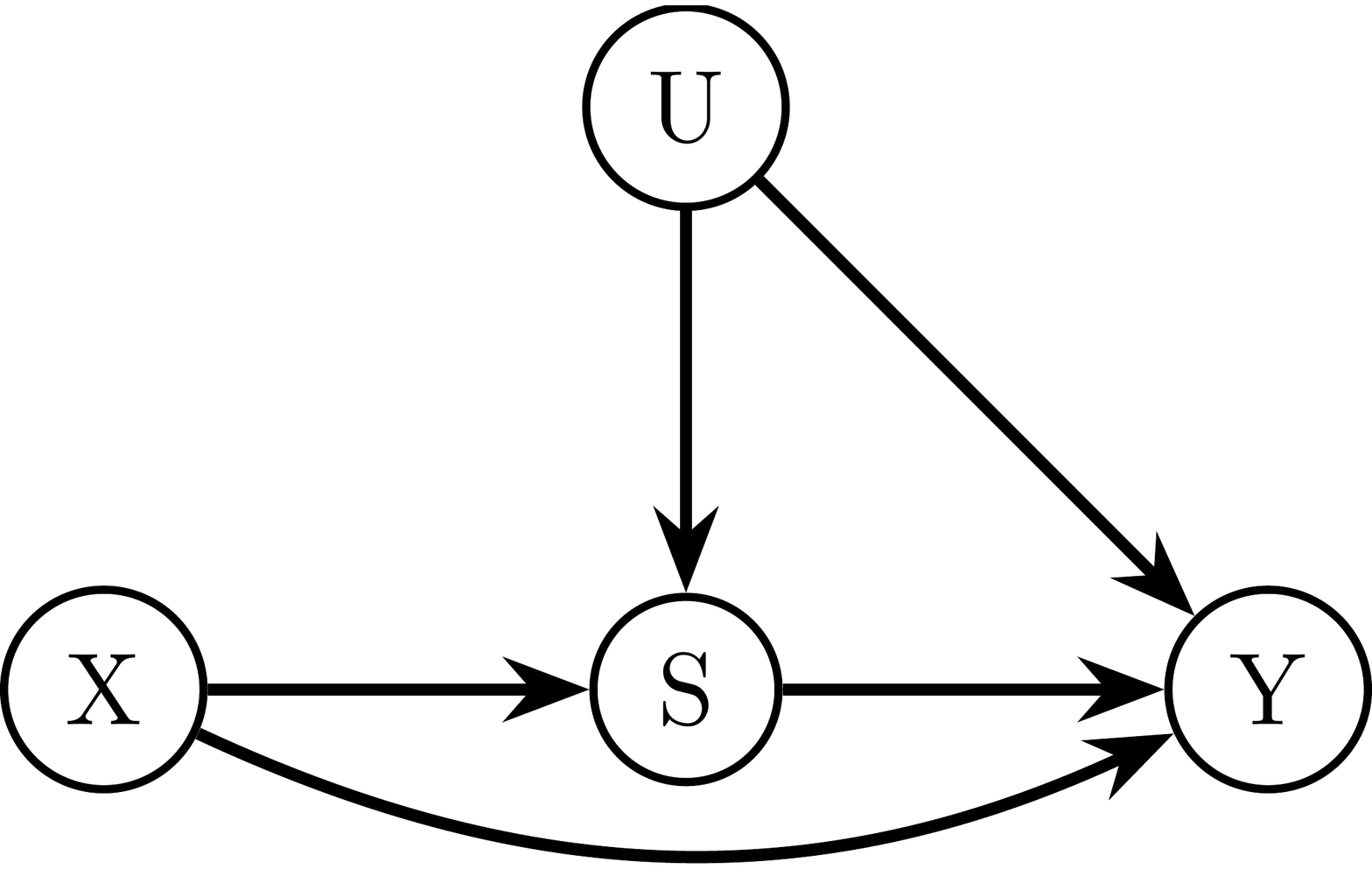}
\caption{Directed acyclic graph (DAG) of a simple scenario which may involve survival bias. $X$ denotes an exposure, $S$ denotes survival until a particular time point, $Y$ is the outcome of interest and $U$ is an unmeasured confounder.}
\label{Figure:DAGbasic}
\end{figure}

Before the modeling framework is formally defined, it is helpful to consider truncation bias in a simple causal graph (Figure \ref{Figure:DAGbasic}).  We are interested in the effect of a binary exposure $X$ (taking values 0 and 1) on our survival outcome $Y$. That is, we aim to assess how the rate of $Y$ at a time $t$ may differ under (hypothetical) interventions on $X$. We observe individuals conditioning on survival until time $t_1$ (Hence the node $S$ in Figure \ref{Figure:DAGbasic}). There is an unknown factor $U$ that also influences the event times. 

The graph provides some immediate insights. The truncation bias due to conditioning on $S$ is described by the non-causal path $X \rightarrow S \leftarrow U \rightarrow Y$. Under the null hypothesis of no causal effect of  $X$ on $Y$, $X\rightarrow Y$ is absent, i.e.\ there is no truncation bias, and standard hypothesis tests are therefore null-calibrated even if we condition on $S$.  Similarly, if $X \rightarrow S$ is absent, truncation bias is neither an issue. This scenario e.g.\ arise if the exposure effect is delayed such that it does not influence entry into the study \citep{vansteelandt2017survivor}. There may also be particular parameterisations of th data generating mechanism that do not lead to bias, which e.g.\ may be derived from additive hazards models \citep{vansteelandt2017survivor}. 

For predictions \textit{per se}, causal effects are not always the primary interests, and the considerations above are not necessarily relevant; indeed, standard models may sometimes be immediately applicable. 

\section{Notation and motivation}
Similar to \citet{vansteelandt2017survivor}, who recently studied survival bias in the additive hazards framework, let $T$ denote lifetimes, $T_0$ indicate the time of left truncation and $C$ denotes the censoring time. Let $X$ be a binary exposure, $L$ is a vector of measured covariates, and $U$ is an unmeasured time-fixed variable which we also denote the frailty, such that $U \bigCI L$, $X \bigCI U$ and $X \bigCI L$. We use superscript notation for counterfactual outcomes such that $T^{X=x}$ denotes the lifetime if, possibly contrary to fact, $X$ is set to $x$. We will assume that event times are generated from the hazard 
\begin{align}
h^{X=x}(t | L,U) = h(t | X=x, L,U) = h_0(t) \exp(\beta_xX + \beta^T_lL) U,
\label{eq: new basic frailty}
\end{align}
where $L$ is a vector of measured covariates and $r(t)=e^{\beta_x}$ is the counterfactual HR under $X$ conditional on $U$ and $L$. Equation \eqref{eq: new basic frailty} is coherent with the classical frailty model, suggested by e.g.\ \citet{vaupel1979impact} and \citet{lancaster1979econometric}. The model in Equation \eqref{eq: new basic frailty} lends on the untestable assumptions that $U$ is time constant, and that $r(t)$ is the same for all levels of $U$. These assumptions allow us to consider $r(t)$ as a marginalised estimand because
\begin{align}
\frac{E_U(h^{X=1}(t |L, U))}{E_U(h^{X=0}(t |  L, U))} = r(t)
\label{eq: marginalised hazard ratio}.
\end{align}
Such ratios are often considered in the statistics and economics literature \textit{individual hazard rates} \citep{hougaard1995frailty, van2001duration, van2016inference}. 

In a deterministic counterfactual framework individual hazards may not be interesting \textit{per se}: At the event time, the individual hazard will be infinity, and at any other time the hazard will be 0, meaning that the individual HR is either 0 or undefined. Nevertheless, $r(t)$ does have a meaning under the data generating mechanism in Equation \eqref{eq: new basic frailty}, which may be interpreted as a particular stochastic counterfactual model: Under these particular assumptions, it may be interpreted as the probability of causation due to $X$ \citep{robins1989probability}, which e.g.\ seems to be used in compensation issues: Assume that a subject exposed to an environmental toxin $X$ experienced an event at $t$, and an insurance company will give compansation based on the probability that the event was caused by $X$. Then, $\frac{r(t)-1}{r(t)}$ may denote the probability of causation. Intuitively, $r(t)$ is the relative effect that an individual cares about: It is the rate ratio under the counterfactual scenarios of being exposed or unexposed. Since $U$ is unknown, $r(t)$ cannot yield precise information about the absolute increase in hazard or risk. 


The frailty model in Equation \ref{eq: new basic frailty} is compatible with a Cox model with hazard ratio $r^{\star}$ of $X$ only if
\begin{align}
 r^{\star}   = \frac{E_U(h(t |X=1,L,  U))}{E_U(h(t | X=0,L, U))}
\label{eq: comparing cox frailty}
\end{align} 
is valid for all $t$. Importantly, under a constant $r(t) = r$, Equation \eqref{eq: comparing cox frailty} will only hold under very special conditions, such as $r=1$ or $\text{VAR}(U)=0$. However, even if the Cox proportional hazard assumption is valid, there is still no simple relation between $r^{\star}$ and $r(t)$ \citep{hernan2010hazards}. In contrast to $r(t)$, it is therefore hard to interpret $r^{\star}$, and it is not a counterfactual rate ratio with an immediate relevance to an individual.

The Cox PH assumption can obviously be assessed from the data during follow-up, which is clearly an advantage of the method. Before the onset of follow-up ($T_0$), however, we have no way to justify the proportional hazards assumption. Hence, not only the frailty model, but also the standard Cox model will rely on untestable assumptions in left truncated samples. We will hereby assume that data are generated from models in which $r(t)=r$ for all $t$, which is analogous to the proportional hazards assumption of the Cox model.


\subsection{Study populations}
Assume that we got a random sample $S_1$ of subjects $1,2,...,n$ from a population $P$ with event times generated by Equation \eqref{eq: new basic frailty}. We observe left truncated lifetimes, i.e. all study subjects got $T > t_0$, where $t_0$ is the time of left truncation. The event times may be censored at $C$ such that for each individual we observe $V=\text{min}(T,C)$, and we assume independent censoring. For the ease of presentation, there are no measured covariates $L$ in the following considerations, but the approach is valid when $L$ is present. Based on the observed event times we aim to estimate $r$. The frailty $U$ is unmeasured, and left truncation is obviously an issue \citep{vansteelandt2017survivor}. 

Assume also that we got summary results from another random sample $S_2$ of $P$, consisting of pairs of individuals sharing the same $U$, and each subject is unexposed to $X$. If $U$ was solely determined by genetic factors, $S_2$ could e.g.\ be a population of monozygotic twin pairs. This sample only contains information on whether each subject survived until $T_0$. In practice, these data could e.g.\ be derived from a twin birth registry. Suppose also that we know the parametric distribution of $U$. This information may allow us to estimate $r(t)$ in a left truncated sample. 

\section{Learning the frailty distribution from the data}
\label{sec: using family}
Let $U$ be a random variable with $\text{E}(U)=1$, i.e.\ any standardised (frailty) distribution with a finite mean. Let $S(t)=\text{P}(T>t)$, i.e. the survival probability unconditional on $U$ and $X$. The cumulative baseline hazard is $H_0(t) = \int_{o}^{t} h_0(t) dt$. Let $T_i$ and $T_j$ denote the time of event for individuals $i$ and $j$ who are monozygotic co-twins. Assuming everybody is unexposed, and let the twin recurrence risk at time $t$ be TRR($t$), which is the relative risk of surviving until time $t$, given a co-twin lived longer than $t$,
\begin{align}
\text{TRR(t)} &= \frac{\text{P}(T^{X=0}_i>t |T^{X=0}_j>t)}{\text{P}(T^{X=0}_i>t)} \nonumber \\
&= \frac{\text{P}(T^{X=0}_i>t, T^{X=0}_j>t )}{\text{P}(T^{X=0}_i>t)^{2}} \quad \text{ because } \text{P}(T^{X=0}_i>t)=\text{P}(T^{X=0}_j>t)\nonumber \\
&= \frac{E[S(t^{X=0}|U)^2]}{S(t^{X=0})^{2}}  \nonumber \\
&= \frac{E[e^{-2H_0(t)U}]}{E[e^{-H_0(t)U}]^{2}} \nonumber \\
&= \frac{L[2H_0(t)]}{L[H_0(t)]^2},
\label{eq:TRR}
\end{align}
where $L$ denotes the Laplace transform of $U$. 


\subsection{Analytic results for the power variance function distributions}
\label{sec:PVFderivations}
Assume that the distribution of $U$ belongs to the large class of power variance function (PVF) distributions, which e.g.\ includes the gamma distribution, the inverse Gaussian distribution, the (compound) and the Poisson distribution. The PVF family also includes the Hougaard distributions\citep{hougaard2012analysis} that are continuous and unimodal and cover the inverse Gaussian distribution as a special case. Similar to \citet{aalen2008survival}, we consider PVF distributions with $\text{E}(U)=1$, and we express the expected value, variance and Laplace transform of the PVF family as
\begin{align} 
\frac{m\rho}{\nu} = 1, \nonumber \\ 
\text{VAR}(U ) & = \frac{(m+1)}{\nu}, \nonumber \\
\qquad L(c|\nu,m) &= e^{-\frac{\nu}{m}(1-(\frac{\nu}{\nu+c})^m)},
\label{generalProperties E1} 
\end{align}
where $\nu > 0$, $m > -1$ and $m\rho > 0$. The survival function under a PVF distribution is
\begin{align}
\label{survivalFunction}
S(t^ {X=0}) &= \text{E}(e^{-H_0(t)U}) \nonumber \\
&=L(H_0(t)|\nu,m) \nonumber \\
&= e^{-\frac{\nu}{m}\left(1-\left(\frac{\nu}{\nu+H_0(t)}\right)^m\right)}.
\end{align}
We will use Equations \eqref{generalProperties E1} and \eqref{survivalFunction} to find $\text{VAR}(U)$ under a particular PVF distribution. 


\subsubsection{Calculations for the gamma distribution}
\label{sec:gammaExample}
To illustrate, we consider the gamma distribution, which is probably the most frequently used distribution in frailty theory. The gamma distribution is mathematically tractable, and it may be theoretically appealing because the heterogeneity of frailty distributions converges to a gamma distribution among survivors \citep{abbring2007unobserved}. However, assuming a gamma distributed $U$ at $t_0$ can typically not be tested, and we must heuristically justify the distribution, e.g.\ because we believe that $U$ consists of a sum of factors, making it continuous in the population.   Using the notation in Equations \eqref{generalProperties E1} and \eqref{survivalFunction}, the gamma distribution arises when $\rho \rightarrow \infty$ and $m \rightarrow 0$ such that $\rho m \rightarrow \eta > 0$, where $\eta$. since $\text{E}(U) = \frac{\eta}{\nu} = 1$, the gamma frailty reduces to a one parameter distribution, and the Laplace transform becomes
\begin{align}
L(c | \nu) = \left(\frac{\nu}{\nu + c}\right)^\nu.
\label{eq gamma laplace}
\end{align}
We may simplify Equation \eqref{survivalFunction} to
\begin{align}
\label{survivalFunctiongamma}
S(t^{X=0}) = \text{E}(e^{-H_0(t)U}) =L(H_0(t) | \nu) = \left(\frac{\nu}{\nu + H_0(t)}\right)^\nu,
\end{align}
and from Equation \eqref{eq gamma laplace} we  simplify TRR(t) to 
\begin{align}
\label{TRRfunctiongamma}
\text{TRR(t)} = \frac{\left(\frac{\nu}{\nu + 2H_0(t)}\right)^\nu}{S(t^{X=0})^2},
\end{align}
We use Equation \eqref{survivalFunctiongamma} to express $H_0(t)$ as a function of $S(t | X=0)$ and $\nu$
\begin{align}
\label{hazardFunctiongamma}
H_0(t) = \frac{\nu(1-S(t^{X=0})^{\frac{1}{\nu}})}{S(t^{X=0})^{\frac{1}{\nu}}}.
\end{align}
Finally, we replace $H_0(t)$ in Equation \eqref{TRRfunctiongamma} by the right side of Equation \eqref{hazardFunctiongamma}, and to find $\nu$ we can numerically solve
\begin{align}
\label{eq: nu solver}
S(t^{X=0})^2 & \text{TRR(t)} - \left(\frac{1}{1+2\left(\frac{1- S(t^{X=0})^{\frac{1}{\nu}}}{ S(t^{X=0})^{\frac{1}{\nu}}}\right)}\right)^{\nu} = 0 \nonumber \\
\end{align}
for any time point $t=t_1$. When $\nu$ is derived, we are able to specify the variance of a gamma distributed $U$. The same logic can be used for other PVF distributions (see the Appendix for R scripts that implement these methods numerically). We consider a counterfactual population of unexposed individuals, and this population is generally unobserved. In practice, we will estimate the $TRR$ from observed quantities. 


In the next section, I will describe how the information on $U$ can be used to obtain estimates of HRs $r$ with a causal interpretation. This requires an estimate of the marginal HR among survivors at a specific time $t_1$, which is approximated by a Cox proportional hazards estimate in an interval $[t_1, t_1 + \delta]$.

\subsection{Using the marginal HR to estimate the causal HR}
\label{sec: observed to causal}
 
We continue to study the PVF family. Assume that our data are generated by a proportional hazards model as in Equation \eqref{eq: new basic frailty}, and let $r_{mar}(t)$ denote the marginal HR among survivors at time $t$. Let $r$ denote the (constant) causal HR conditional on $U$, such that \citep{aalen2008survival}
\begin{align}
r_{mar}(t) = r \Bigg( \frac{1+\frac{H_0(t)}{\nu}}{1+\frac{rH_0(t)}{\nu}} \Bigg)^{m+1},
\label{equation:observedHazardExpressedWithCausal}
\end{align}
which means that
\begin{align}
\label{eq:observedRiskSolver}
r_{mar}(t)^{\frac{1}{m+1}} \left(1+\frac{H_0(t)r}{\nu}\right) = r^{\frac{1}{m+1}}\left(1+\frac{H_0(t)}{\nu}\right).
\end{align}
Assume that $\nu$ is derived by the approach in Section \ref{sec:PVFderivations}, e.g. using twin data. The marginal HR at a particular time $t_1$ is approximately derived from a Cox proportional hazards model in an interval $[t_1, t_1 + \delta]$. For the gamma distribution, $m\rightarrow 0$, and we find $r$ analytically by solving
\begin{align}
 r = r_{mar}(t) \left(1+\frac{H_0(t)}{\nu} - \frac{H_0(t)r_{mar}(t) }{\nu}\right)^{-1}
\label{eq:gammaRiskSolver} 
\end{align}
for any time point $t=t_1$, where we use Equation \eqref{hazardFunctiongamma} to find $H_0(t)$. For the inverse Gaussian distribution, $m=-0.5$, and we obtain $r$ analytically by solving a quadratic equation
\begin{align*}
r^{2}\left(1+\frac{H_0(t)}{\nu}\right) - r\left(r_{mar}(t)^{2}\frac{H_0(t)}{\nu}\right) - r_{mar}(t)^{2} = 0. \\
\end{align*}
For other distributions in the PVF class, we may solve Equation \eqref{eq:observedRiskSolver} with respect to $r$ numerically. Confidence intervals of $r$ can be found numerically, as suggested in Appendix \ref{appendix: confidence intervals}. In Appendix \ref{appendix: simulations to verify the approach}, a simulation study of plausible scenarios was performed, in which the adjusted 95\% confidence intervals obtained approximately $95\%$ coverage of the true $r$ in all scenarios. 




In applied settings, we may expect a slight bias towards a null effect because because the marginal $HR$ at $t_1$ is approximated by a Cox proportional hazards estimate in $[t_1, t_1 + \delta]$, as suggested in Section \ref{sec:gammaExample}. This means that we adjust for the impact of $U$ until $t_1$, but we do not adjust for the effect of $U$ during follow up $[t_1, t_1 + \delta]$. In the simulations in Appendix \ref{appendix: simulations to verify the approach}, $t_1=50$ was much larger than $\delta =2$, and the bias was negligible. 

Nevertheless, \citet{mcnamee2017serious} has recently suggested a heuristic approach to deal with this model misspecification if the frailty is gamma distributed: Assume that we follow a population from $t_0$, and let $t_{median}$ be the median of all event times in the population. Under a gamma distributed frailty, replacing $t$ by $t_{median}$ in Equation \eqref{eq:gammaRiskSolver}, will yield adequate estimates of $r$. This approach was shown to perform satisfactory in simulations \citep{mcnamee2017serious}. Hence, if we follow subjects from $t_0$, we may plug in frailty estimates from twin data into the expression suggested by \citet{mcnamee2017serious}, and thereby explore the truncation bias, e.g.\ in RCTs.  




\subsection{When is survival bias an issue?} 
The magnitude of survival bias varies with (i) time, (ii) the size of the observed TRR and (iii) the parameterisation of the frailty distribution. We shall consider some scenarios, using the derivations in Sections \ref{sec: using family}-\ref{sec: observed to causal}. 

\begin{figure}[ptb]%
\centering
\includegraphics[width=9cm]{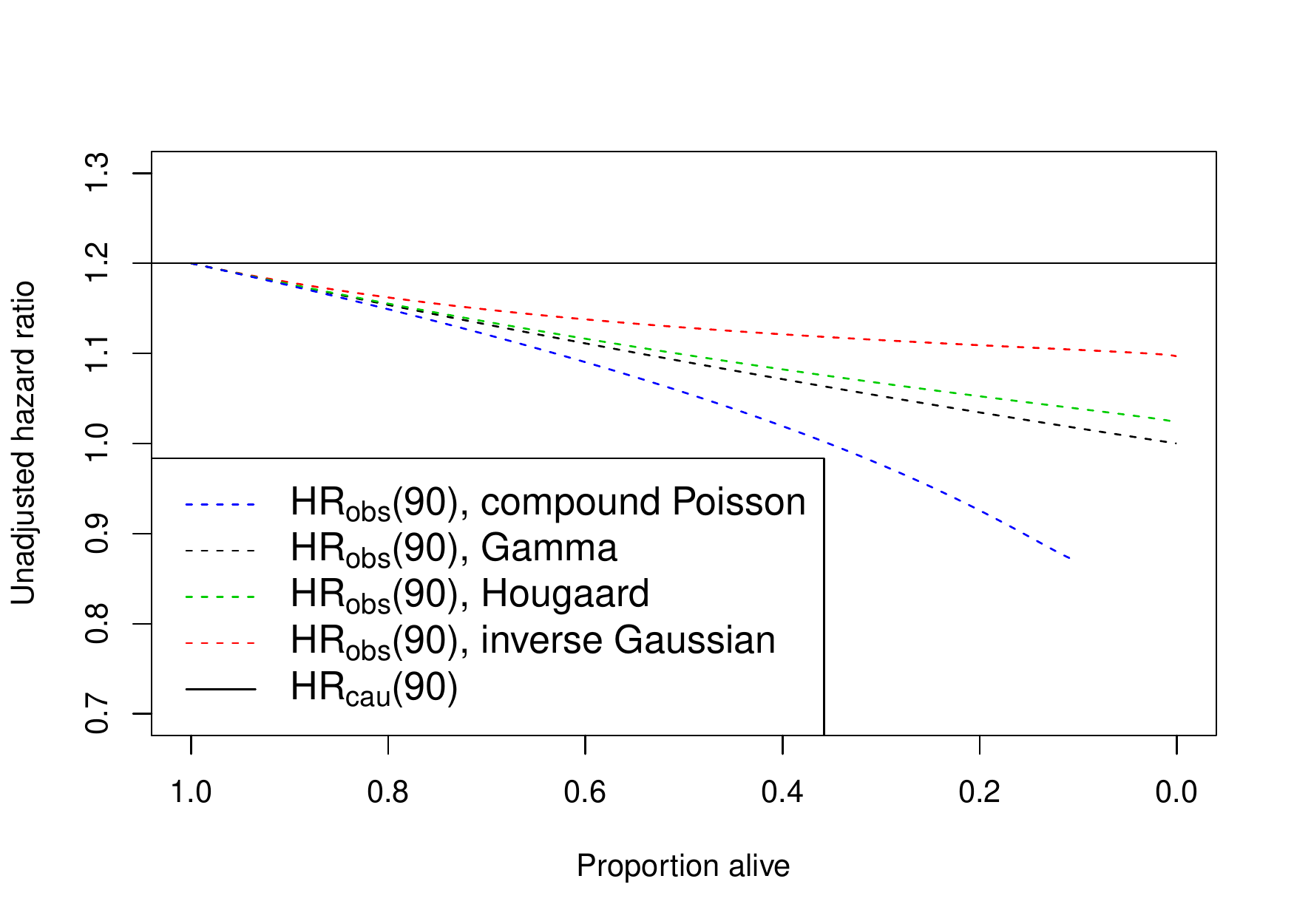}
\caption{Hazard ratios plotted as a function of the proportion alive at $t_0$. The casual hazard ratio is $HR_{cau}(t)=1.2$ (solid line). The dashed lines show $\text{HR}_{obs}(t=t_1)$, where assuming different frailty distributions: Gamma distribution (black), inverse Gaussian distribution (red), Hougaard distribution (green) and compound Poisson distribution (blue).}
\label{figure: TRR over time}%
\end{figure}

First, Equation \eqref{equation:observedHazardExpressedWithCausal} shows that the bias increases with $t$. In particular, when $t \rightarrow \infty$, we have that $r_{mar}(t) \rightarrow r^{-m}$. For the gamma distribution, $m \rightarrow 0$, and $r_{mar}(t) \rightarrow 1$. Hence, the marginal HR will be attenuated towards a null-effect \citep{van2001duration}. For some conditions, it may be that only a fraction of the population is susceptible, e.g. for a particular disease. In these scenarios, the compound Poisson model is convenient, because it allows for a non-susceptible fraction. For the compound Poisson distribution, $m > 0$, and 
\begin{align*}
\lim_{t\to\infty} r_{mar}(t) = r^{-m} < 1 \qquad \text{ if } r > 1 \nonumber \\
\lim_{t\to\infty} r_{mar}(t) = r^{-m} > 1 \qquad \text{ if } r < 1,
\end{align*}
which means that we eventually will observe $r_{mar}$ in the opposite direction of $r$. This may have important implications: The marginal HR is not only a biased estimate of $r$, but it may also be invalid for hypothesis testing of an effect of $r$. Even though this relation is well-known in the methodological literature \citep{robins1989probability,van2001duration}, it may be under-appreciated in the applied literature \citep{burgess2015commentary}. In Figure \ref{figure: TRR over time}, the relation between the unconditional HR and left truncation is displayed for some PVF distributions with variance equal to 1. In all scenarios the conditional HR is 1.2, but the unadjusted HR declines a function of the population fraction lost to left truncation. 

\begin{figure}[ptb]%
\centering
\includegraphics[width=9cm]{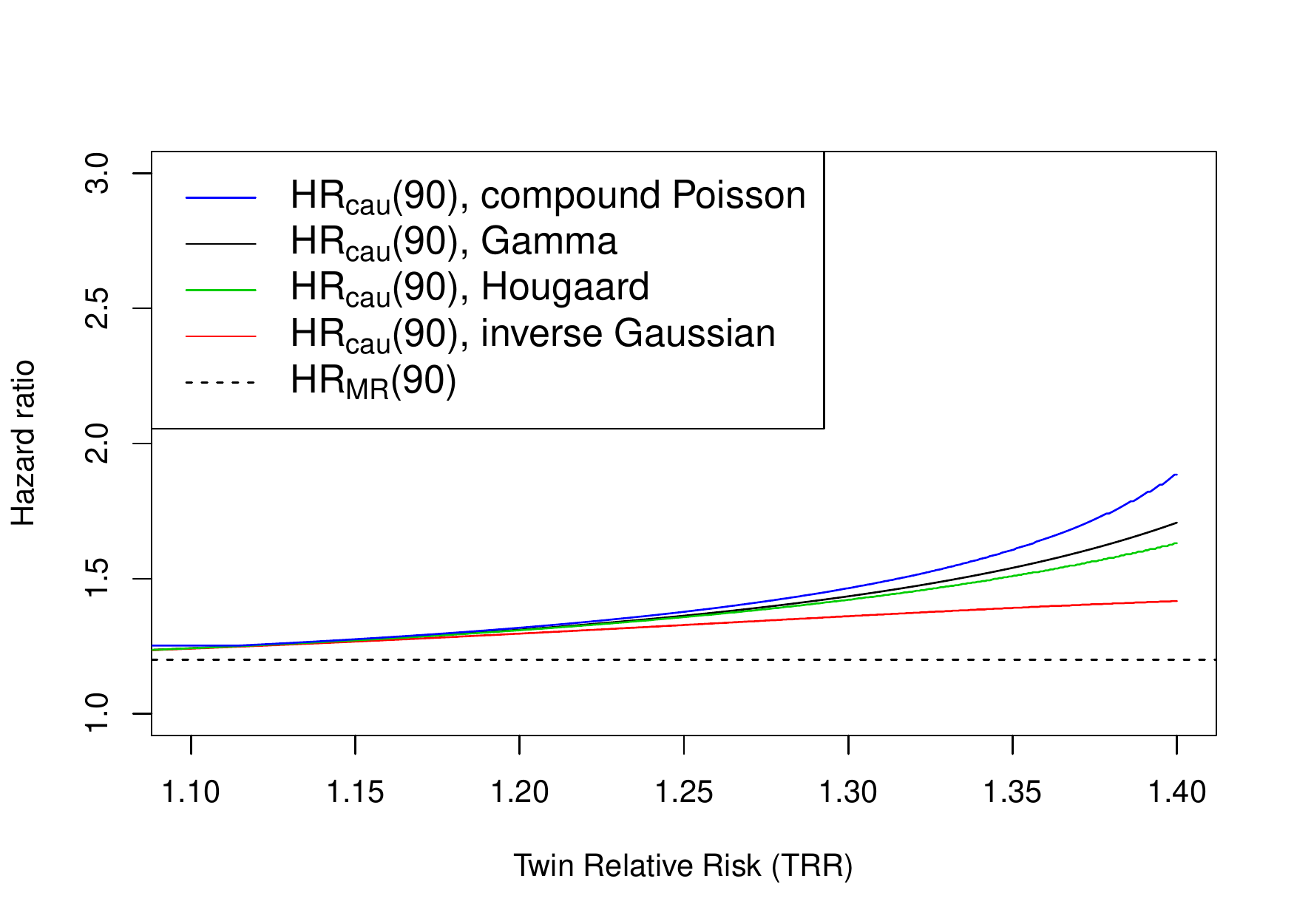}
\caption{Hazard ratios plotted as a function of the TRR. We have assumed that the marginal hazard ratio conditional on survival is $r_{mar}(t=t_1)=1.2$ (dashed line). The solid lines show $\text{HR}_{cau}(t=t_1)$, i.e. the causal hazard ratio conditioning on $U$, assuming a gamma distribution (black), an inverse Gaussian distribution (red), a Hougaard distribution (green) and a compound Poisson distribution (blue).}
\label{figure: TRR bias}%
\end{figure}

Second, a large TRR$(t)$ yields a large variation in risk between individuals \citep{valberg2017surprising}. Intuitively, we may also think that a large variation in risk leads to larger survival bias. To perform a numerical evaluation of the bias, let the population be assessed at time $t_1$ such that $S(t_1)=0.9$. In Figure \ref{figure: TRR bias}, we display $r$ as a function of the TRR$(t_1)$ when $r_{mar}(t_1)= 1.2$. For a fixed $r_{mar}(t_1)$, the conditional $r$ increases TRR$(t_1)$. 

\section{Extension to IV estimates}
\label{sec: extension to IV estimates}
Instrumental variable (IV) approaches may be useful to identify causal relations. To find the effect of an exposure on an outcome, these techniques rely on an additional variable, the instrument. Let $L$ be measured covariates, and let $U$ be a possibly unmeasured confounder. Given $L$, an instrument must satisfy the following assumptions to obtain unbiased effect estimates:
\begin{itemize}
\item The instrument $G$ is associated with the exposure $A$, i.e.\ $G \nbigCI A | L$.
\item The only path leading from $G$ to the outcome $Y$ goes through the exposure, $ G \bigCI Y | A, U, L $. 
\item $G$ is independent of any unmeasured factor $U$ that confounds the exposure-outcome relation $ G \bigCI U | L $. 
\end{itemize}
For more information on the IV assumptions, see e.g. \citet{didelez2007mendelian}. In the biomedical literature, the number of analyses based on IVs are increasing. Mendelian Randomisation (MR) studies are particularly popular, and such analyses rely on genetic variants as instruments. There is, however, a temporal aspect of MR studies. The genetic variants are allocated at conception ($t=0$), but the follow-up starts in adult life. Hence, survival bias is potentially a major issue because (i) there is often a large time-lag between perceived randomization at $t_0$ and follow-up, and (ii) MR holds the promise to reveal causal effects. 

\begin{figure}
\centering
\includegraphics[width=4.5cm]{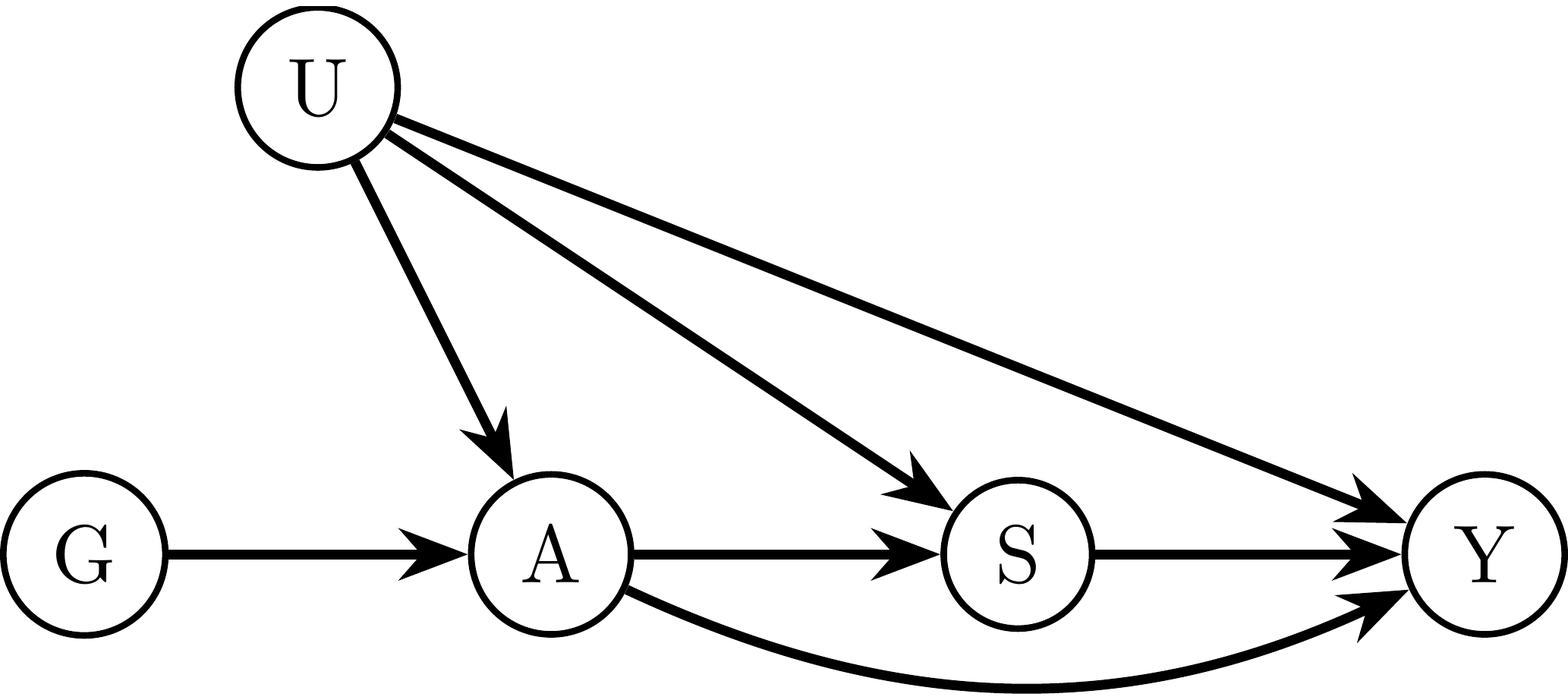}
\caption{Causal graph of a Mendelian Randomisation study with loss to follow-up. Here, $G$ is the genetic instrument, $A$ denotes an exposure, $S$ denotes survival until a particular time point, $Y$ is the outcome of interest and $U$ is an unmeasured confounder.}
\label{Figure:DAGsurvival}
\end{figure}

A causal structure of an IV setting with survival outcomes is shown in Figure \ref{Figure:DAGsurvival}. Here, $S$ is survival until time $t_1$. Until now we have considered a simple binary exposure, but we will hereby let the exposure $A(t)$ be a time-varying continuous variable expressed as a  function of the binary instrument $G$ and the unknown component $U$. Let the data generating mechanism of $A(t)$ be
\begin{equation}
A(t) = b_0(t) + b_g G + b^T_l L + f(U) + e(t),
%
%
\label{eq: initial A and G}
\end{equation}
where $f(U)$ is a random variable such that $E( f (U | G)) = E( f (U ))$, $L$ denotes the measured covariates assuming that $L \bigCI  G$ and $L \bigCI  U$, and $e(t)$ is a residual error independent of $U$ which we do not specify further. If $\text{E}(f(U))=z$, we substitute $b_0(t)$ with $b_0^*(t) = b_0(t) + z$ and $f(U)$ by $f^*(U) = f(U)-z$, such that $\text{E}(f^*(U))=0$. In Equation \eqref{eq: initial A and G} we assume that $b_g$ is time constant for each subject, which is not trivial to 
justify in practice (see e.g.\ \citet{abbring2007econometric} who discuss this in an economics context). 

\citet{tchetgen2015instrumental} suggested a proportional hazards strategy for IV analyses, which is only valid for rare outcomes due to the non-collapsibility of the hazard ratio. By modelling the relation between $U$ and survivial ($T$), I will suggest a proportional hazards approach that is not only valid under the rare events assumptions. Similar to \citet{tchetgen2015instrumental}, I will consider a proportional hazards model for the outcome $Y$,
\begin{align}
h(t | U, A(t), G) &= h(t | U, A(t)) \nonumber \\ 
&= h_0(t) e^{\beta_aA(t) + \beta^T_l L} \psi(U),
\label{Assumptions equality 1}
\end{align}
where $\psi(U)$ is a function of $U$ independent of $A$. In Equation \eqref{Assumptions equality 1}, the first line is justified by the graph in Figure \ref{Figure:DAGsurvival}, and the second line displays the assumed causal hazard function. Hence, $\beta_a$ denotes the causal effect of $A(t)$ on $Y$ conditional on $U$, which is our estimand. Different from \citet{tchetgen2015instrumental}, we allow $A(t)$ be time-varying, but we restrict $\psi(U)$ to be time-constant.  


To estimate $\beta_a$, we must rely on information about $G$ and $U$, and we will use the results derived in Section \ref{sec:PVFderivations}. It may indeed be unappealing that this IV analysis will require explicit assumptions about $U$, because the motivation for using an instrument is to avoid modeling $U$. Nevertheless, in the Mendelian randomization context, one may also argue that the IV assumptions are not met, because $G \nbigCI U$ under left truncation. To deal with this issue, I will model the relation between $U$ and survival $T$, but I will still be agnostic about the relation $U \rightarrow A$.


\subsection{Causal estimates in Mendelian randomisation studies.}
We aim to estimate $\beta_a$. Writing out Equation \eqref{Assumptions equality 1} gives
\begin{align}
h(t |  U, A(t), G) &=  h_0(t) e^{\beta_a(b_0(t) + b_g g + b_v V +  f(U) + e(t)) + \beta^T_l L} \psi(U).
\label{Causal hazard 1 g}
\end{align}
We introduce the parameter $U^{\star}$ such that $U^{\star}=e^{\beta_af(U)}\psi(U)$. Similar to Section \ref{sec:PVFderivations}, we assume that $U^{\star}$ has a distribution with finite mean in the population, and we assume that $U^{\star}$ is standardised such that $\text{E}(U^{\star})=1$.  Inserting $U^{\star}$ in Equation \eqref{Causal hazard 1 g}, we obtain
\begin{align}
h(t |  U, A(t), G) 
&= h_0(t) e^{\beta_a(b_0(t) + b_v V + b_g g + e(t))+ \beta^T_l L}U^{\star}.
\label{IvHazard}
\end{align}
Let $g_1$ and $g_2$ be two variants of the instrument $G$. We can then consider the genetic HR

\begin{align}
\text{HR}_{G} &= \frac{h(t | U^{\star}, G=g_1) }{h(t | U^{\star},G=g_2)} \nonumber \\
&= e^{\beta_ab_g(g_1 - g_2)}. \nonumber \\
\label{expression:analogueToStd}
\end{align}
We can find an estimate $\hat{b}_{g}$ of $b_{g}$, e.g.\ by fitting a linear regression based on Equation \eqref{eq: initial A and G}. Furthermore, $\text{HR}_{G}$ can be estimated with the strategy in Section \ref{sec: using family} such that 
\begin{align}
\hat{\beta}_a = \frac{\log[\hat{\text{HR}}_{G}]}{\hat{b}_g(g_1 - g_2)}.
\label{expression:identification}
\end{align} 
In contrast to \citet{tchetgen2015instrumental}, in Equation \eqref{expression:identification} it is not assumed that $S(t | U, A(t), G) \approx 1$. \citet{tchetgen2015instrumental} required the rare events assumption, because they used standard Cox proportional hazards regression to find the marginal HR under a data generating mechanism similar to Equation \eqref{Assumptions equality 1}. Hence, they assumed that the causal effect of the exposure conditional on $U$ is proportional on the hazard scale, but they fitted a Cox proportional hazards model unconditional on $U$. Due to the non-collapsibility of the HR, the unconditional model is approximately correct only if the event is rare. In this article, we do use the estimates from a (mis-specified) marginal proportional hazards model in an interval $[t_1,t_1 + \delta]$ as an approximation to $r_{mar}(t_1)$. However, we do not require that the true, marginal HR can be correctly estimated by a Cox model at other times $t$, and we therefore do not need the rare events assumption. 

Until now the instrument $G$ has been considered to be binary. A binary $G$ may be reasonable when the instrument is a single gene, which initially was the standard approach for MR analyses. Today, however, most analysts use instruments that are combinations of multiple variants, usually quantified in a continuous genetic risk score \citep{burgess2013use}. This approach will often increase the power of the study, but it also requires all the single variants to satisfy the IV assumptions. Our approach is readily applicable to continuous instruments. Specifically, we then consider $G$ to be a continuous variable, and Equation \eqref{eq: initial A and G} thereby require $G$ to have an additive effect on the Exposure $A(t)$. The remaining derivations in Section \ref{sec: extension to IV estimates} will all be valid for both a continuous and a binary $G$. 

\section{An illustrative example}
\label{sec: full example section}
\subsection{The effect of alcohol on all-cause mortality} 
\label{sec: standard alcohol example}
\citet{almeida2015excessive} assessed the impact of alcohol consumption on all-cause mortality in a sample with 3496 old men (aged 70-89 years at baseline). To do this, they used genetic information on the Alcohol Dehydrogenase 1B (ADH1B) gene. A mutation in the gene, which was carried by 225 of the subjects, leads to abnormal metabolism of ethanol, and carriers experience an unpleasant flushing while drinking. It is well-known that carriers consume less alcohol \citep{holmes2014association}, and their reduced consumption is thought to be independent of confounding factors. 

Following \citet{almeida2015excessive}, we will consider the effect of (ADH1B) gene on all-cause mortality, and not the effect of alcohol per se.  At baseline there should not be confounders that affects all-cause mortality and carrying ADH1B. However, in a left truncated sample the mortality HR between carriers and non-carriers of the genetic variant ADH1B (now, considered to be the exposure $X$ in Figure \ref{Figure:DAGbasic}) is not easy to interpret; at least it cannot be interpreted as the counterfactual rate ratio of carrying vs not carrying the ADH1B in a subject. To illustrate, we consider the marginal HR $$r_{Almeida} = 0.68 \text{ }[0.54,0.87]$$ in carriers of the mutation, which was derived by \citet{almeida2014alcohol}. With the frailty methods, we may explore how this marginal estimate may deviate from a counterfactual rate ratio under explicitly defined model assumptions. 

First, we consider the age of 75 years, which is approximately the mean age at baseline in the study by \citet{almeida2014alcohol}, such that $r_{mar}(75) = r_{Almeida}$. Analyses of a Scandinavian registry of 9272 male twins suggest that the relative probability of surviving 75 years, given a co-twin survived 75 years, is 1.27 in men \citep{iachine2006genetic}. Therefore we let $\hat{\text{TRR}}(75)=1.27$, assuming that the probability of surviving until 75 years is shared among monozygotic twins. Here, we should ideally have assessed the TRR(75) in a (counterfactual) population of non-carriers of the ADH1B mutation, but we used the whole population as a crude estimate (in Scandinavians less than 4\% carry the mutation, which is even rarer than in Australians \citep{linneberg2010genetic}). 

According to Australian cohort life tables of subjects, 0.56 of Australian men born 1931-1936 survive until age 75 years \citep{rowland1997demography}. We assume a gamma distribution of $U$; heuristically we then believe that everybody has a hazard larger than 0 for dying at any time, and the individual hazards vary continously in the population. Intuitively, these hazards will arise due to a combination of several unmeasured factors denoted $U$. We will use the values $\text{TRR(75)}=1.27 \text{ }[1.20,1.34]$, $S(75)=0.56$ and $r_{mar}(75)=0.68 \text{ }[0.54,0.87]$ to estimate the causal HR conditional on $U$. To highlight the magnitude of bias due to $U$, we first assume that $\text{TRR(75)}$ and $S(75)$ are true values without uncertainty, and we numerically solve Equation \eqref{eq: nu solver} to find $\nu=0.846$. Then, we use Equation \eqref{eq:gammaRiskSolver} to estimate the causal HR by 
\begin{align*}
 \hat{r} = r_{mar}(75) \Big(1+\frac{H_0(75)}{0.846} - \frac{H_0(75)r_{mar}(75) }{0.846}\Big)^{-1} = 0.52 \quad [0.37, 0.77].
\label{eq:gammaRiskSolver2} 
\end{align*}
 The 95\% confidence interval is simply derived by plugging in the confidence limits of $r_{mar}(75)$ into Equation \eqref{eq:gammaRiskSolver}, and these estimates are valid due to the monotonic relation between $r$ and $r_{mar}$. We thereby assume that $TRR(75)$ and $S(75)$ are the true values in the population. 

The frailty analysis suggests that $r_{mar}(75)$ is a conservative estimate of $r$. To derive this estimate, we have required that a weighted average of $r(t)$ before follow up is equal to the the causal HR during follow up. Allthough this is a straight-forward extension of the standard Cox proportional hazards assumption, it is an untestable assumption, and interpreting $\hat{r}$ causally requires very strong structural assumptions. Rather, $\hat{r}$ gives us an impression on how the crude HRs from Cox models may differ substantially from HRs that actually have a causal interpretation.   

We may also obtain estimates of $r$ by assuming other parameterisations of $U$. In Table \ref{tab: different distributions example}, we have shown results for 4 PVF distributions, using plug-in confidence intervals ($95\% \text{ CI plug-in}$). In this example, the results are robust to the parameterisation of $U$. Intuitively, this is due to the relatively small $TRR(t_1)$. However, this robustness does not necessarily apply to other scenarios \citep{stensrud2017exploring}, e.g. for larger $TRR(t_1)$, as also seen in Figure \ref{figure: TRR bias}.

In applications, it is not sufficient to account for the uncertainty in $r_{mar}$. We must also consider the uncertainty in the other summary estimates that are used, i.e.\ $\text{TRR}(t_1)$ and $S(t_1)$. A numeric approach to derive such confidence intervals is described in Appendix \ref{appendix: confidence intervals}. We use this approach to account for the uncertainty in $\text{TRR(75)}=1.27 \text{ } [1.20, 1.34]$ in the 4 frailty distributions, and these results are displayed in Table \ref{tab: different distributions example} ($95\% \text{ CI numeric}$). In this example, these estimates are similar to the plug-in estimates. 

\begin{table}[ht]
\centering
\begin{tabular}{lrrrr}
Distribution of $U$ & $\hat{r}_{[t_1, t_1 + \delta]}$ & $95\% \text{ CI plug-in}$  & $95\% \text{ CI numeric}$ \\ 
  \hline
Gamma & 0.52 & [0.37, 0.77] & [0.35, 0.74]\\ 
  Inverse Gaussian & 0.53 & [0.37, 0.79] & [0.36, 0.77]\\ 
  Hougaard ($m = -0.125$) & 0.52 & [0.37, 0.77]  & [0.35, 0.77]\\ 
  Compound Poisson (10\% nonsusceptible) & 0.52 & [0.38, 0.77] & [0.36, 0.76]\\  
\end{tabular}
\caption{Estimates of the causal hazard ratio of the ADH1B variant on all-cause mortality, conditional on the frailty. Results from 4 different PVF distributions are displayed.}
\label{tab: different distributions example}
\end{table}



\section{Discussion}
Conventional HRs are difficult to interpret \citep{stensrud2017exploring, aalen2015does, hernan2010hazards,robins1989probability}. This article explores the causal understanding of HRs. By modelling the unobserved heterogeneity in disease risk, conventional hazard ratios are compared to conditional HRs with a causal interpretation. This approach may be useful for sensitivity analysis, e.g.\ to explore how a HR from Cox model may approximately be relevant to individual patients.

More generally, this article highlights a link between frailty models in survival analysis and causal inference \citep{stensrud2017exploring}: Interpreting estimates from Cox regressions may be challenging, e.g. due to non-collapsibility \citep{martinussen2013collapsibility} and left truncation. However, by using frailty models with strong parametric assumptions, we find estimates that are easier to interpret: We identify the effect of the exposure conditional on the unobserved variable $U$, and intuitively this is the effect of the exposure on the individual level.

When measured covariates are able to explain most of the heterogeneity in risk, the frailty approach may be less desirable than alternative approaches, in which covariates are included in the model. For many conditions, however, unmeasured factors may have considerable impact. In such situations, the frailty approach seems to be particularly attractive, e.g.\ in sensitivity analyses.

I have suggested an approach that only relies on summary data of the risk heterogeneity, e.g.\ derived from twin registries. This is desirable, because individual level data are often unavailable. However, If we got access to individual level data, it may be better to perform a joint analysis (see e.g. \citet{van2001duration}).  

By using twin data, I attempt to adjust for heterogeneity due to (unmeasured) genetic factors and shared environment. Using data from monozygotic twins is desirable, because such co-twins are genetically identical and expected to share several environmental factors. Nevertheless, non-shared environmental factors will not be captured unless they are included as measured covariates $L$. If such factors are unmeasured and have large (multiplicative) effect on the hazard function, we may underestimate the magnitude of survival bias. Furthermore, assuming a time-constant $U$ is simplistic, in particular because co-twins may influence each other over time, e.g.\ in health seeking behaviour. We have also implicitly assumed that twins are representative of the general population. In particular, unmeasured factors that influence survival until follow-up at $t_1$ should be similarly distributed in twins as in the general population. Recently, it has been suggested that monozygotic twins live longer than the general population, but the difference was found to be modest \citep{sharrow2016twin}. 

This article has considered proportional hazards models with time to death as outcome. The approach can also be used to other time to event outcomes. e.g.\ time to progression of a disease. For many diseases we will expect the survival bias to be larger \citep{stensrud2017exploring}; we have shown that a larger variance in the unobserved heterogeneity yields a larger survival bias, and the familial clustering of several diseases is considerably larger than the familial clustering of longevity. For such outcomes, we may deal with survival bias by introducing correlated unknown factors ($U_1$ and $U_2$) that influence the time-to-event, respectively, and thereafter use theory for separate, correlated frailty variables (see e.g.\ \citet{aalen2008survival} chapter 6.6). 

The survival bias is a particular issue in MR studies, due to the long time lag between randomisation and follow up \citep{boef2015mendelian,martinussen2017instrumental}. Interpreting IV estimates for time-to-event outcomes is not straightforward \citep{burgess2015commentary}, but such methods have recently been developed, under explicitly defined assumptions. In particular, estimates from the Aalen additive model do not suffer from survival bias under some explicit parameterisations of $U$ \citep{tchetgen2015instrumental, martinussen2016instrumental, li2015instrumental}. \citet{mackenzie2014using} considered a Cox proportional hazards model with additive unobserved confounding, but it relies on some unrealistic assumptions \citep{tchetgen2015instrumental}. Very recently, estimation under a IV structural Cox model for the treatment effect of the treated was suggested \citep{martinussen2017instrumental}. This approach allows us to estimate a causal HR when subjects are follow from $t_0$, but it may still be hard to handle left truncated samples.

\section*{Acknowledgements}
I would like to thank Odd O. Aalen, Kjetil R\o ysland, Morten Valberg and Susanne Strohmaier for their support and discussions on this manuscript. 

\clearpage
\bibliography{timeIV.bib}
\bibliographystyle{biorefs}

\clearpage

\appendix
\setcounter{table}{0} \renewcommand{\thetable}{B.\arabic{table}}

\section{Derivation of confidence intervals}
\label{appendix: confidence intervals}
To derive estimates of the causal hazard ratio $r_{cau}$, we need to consider three estimates with uncertainty: 1) The marginal HR ($r_{mar}$), 2) The  familial risk (twin recurrence risk) at time $t_1$ ($TRR(t_1)$), and 3) The probability of surviving until $t_1$ ($S(t_1)$). We assume that these estimates are derived from independent sources. Then, it is simple to derive confidence intervals of $HR_{cau}$, using a numeric evaluation. Since $\hat{r}_{mar}$ is derived from a Cox model, we have that $\log(\hat{r}_{mar})$ is approximately normally distributed. Similarly, $\hat{TRR(t_1)}$ is estimated as a relative risk, and hence $\log(\hat{TRR(t_1)})$ is approximately normally distributed. By considering survival until $t_1$ as a sample proportion, $\hat{S}(t_1)$ is approximately normally distributed for large $n$ without transformation. We can use this to find a numeric confidence interval, based on $n$ simulated estimates 
\begin{enumerate}
\item Let the estimate of $r_{mar}$ be $\hat{r}_{mar}$ with 95\% confidence interval $[\hat{r}_{mar, min}, \hat{r}_{mar, max}]$. Then,  
$$
\log(\hat{r}_{mar}) \dot\sim N\left(\log(\hat{r}_{mar}), 
\frac{\log(\hat{r}_{mar}) - \log(\hat{r}_{mar,min})}{\Phi^{-1}(0.975)}\right),
\label{eq: simulation step 1}
$$
where $\Phi^{-1}(c)$ is the inverse cumulative standard normal distribution. Hence, for each $i$ in $1,2,...,n$ we draw $log(\hat{r}_{mar, i})$ from this distribution.
\item Let the estimate of $TRR(t_1)$ be $\hat{TRR(t_1)} \text{ 95\% CI: } [\hat{TRR(t_1)}_{min}, \hat{TRR(t_1)}_{max}]$. Then,  
$$
\log(\hat{TRR(t_1)}) \dot\sim N\left(\log(\hat{TRR(t_1)}), \frac{\log(\hat{TRR(t_1)}) - \log(\hat{TRR(t_1)}_{min})}{\Phi^{-1}(0.975)}\right).
\label{eq: simulation step 2}
$$ 
Hence, for each $i$ in $1,2,...,n$ we draw $log(\hat{TRR(t_1)}_{i})$ from this distribution.
\item Let the estimate of $S(t_1)$ be $\hat{S}(t_1) \text{ 95\% CI: } [\hat{S}(t_1)_{min}, \hat{S}(t_1)_{max}]$, e.g.\ derived from Greenwood's formula. Since the 95\% CI is on the form $\hat{S}(t_1) \pm z_{\alpha/2} \text{se} (\hat{S}(t_1)) $  We draw $\hat{S}(t_1)_{i} \sim N\left(\hat{S}(t_1), \frac{\hat{S}(t_1) - \hat{S}(t_1)_{min}}{\Phi^{-1}(0.975)}\right)$.
\item Finally, for each $i$ in $1,2,...,n$ we obtain an estimate $r_{cau, i}$ by plugging the estimates in 1.-3. into the algorithm described in the main text. A $(1-\alpha)$ confidence interval of $r_{cau, i}$ is found by using the $\frac{\alpha}{2}$ and $\frac{1-\alpha}{2}$ quantiles of the estimates. 
\end{enumerate}

\section{Simulations to verify the approach}
\label{appendix: simulations to verify the approach}
A simulation algorithm was derived to check the validity of the causal estimates in Section \ref{sec:PVFderivations}. This algorithm consists of 3 steps, and the R code to implement the approach is attached. Below is a brief description of the approach in natural language.  
\subsubsection*{Step 1. Simulate survival times and find the marginal HR.}
\begin{itemize}
\item Select $n_1$ unexposed subjects. Each $i$ in $1,2,...,n_1$ is characterised by an individual \textit{frailty} $U_i$, and we randomly draw the variable $U_i \sim \text{Gamma}(\text{mean} = 1,\text{VAR}=\nu)$.
\item Similarily, we select $n_2$ exposed individuals. For each $j$ in $1,2,...,n_2$ we randomly draw $U_j \sim \text{Gamma}(1,\nu)$.
\item Let the causal hazard ratio of the exposure $X$ conditional on $U$ be $r$.
\item Assume that the baseline hazard rate is constant, $h_0(t)=h$. To obtain survival times for each $i$, we first draw $W_i \sim \text{Uniform}[0,1]$. Then we find the survival time  $$S_i = \frac{\log(W)}{h U_i},$$ as suggested by \citet{bender2005generating}.
\item Similarly, for each $j$ we draw $W_j \sim \text{Uniform}[0,1]$, and we find the survival time $$S_j = \frac{\log(W)}{r h U_j}. $$
\item Assume follow up starts at $t=t_1$.
\item We then select the sets of individuals $L_1$ such that $i \in L_i$ iff $S_i > t_1$, and $L_2$ such that $j \in L_j$ iff $S_j > t_1$.
\item To derive the marginal HR, we fit a Cox proportional hazards model to the subjects in $L_1$ and $L_2$, using exposure $X$ as the only covariate.
\end{itemize}
\subsubsection*{Step 2. Simulate twin recurrence risks.}
\begin{itemize}
\item To estimate the twin recurrence risk, select $n_3$ unexposed twin pairs. This sample is independent of the sample in Step 1. 
\item For each twin pair $m$ in $1,2,...,n_3$, we randomly draw the shared $U_m \sim \text{Gamma}(1,\nu)$.
\item For each twin pair, we selected a random co-twin $m_1$. We then find the overall risk of surviving until $t_1$, $$ S(t^{X=0} > t_1) = \frac{1}{n_3}\sum_1^{n_3} I(S_{m_1} > t_1), $$ where $I(S_i > t_1)$ is an indicator function taking value 1 if $S_i > t_1$ and 0 otherwise.
\item Let $L_{3}$ denote the set of survivors such that $m_1 \in L_3$ iff $S_{m_1} > t_1$. Denote the number of subjects in $L_{3}$ as $n_{L_{3}}$. Then, we find the conditional probability of survival in $m_2$ given $m_1$ survived $$ S(t_{m_2}^{X=0} > t_1 | t_{m_1}^{X=0} > t_1)  = \frac{1}{n_{L_{3}}}\sum_1^{n_{L_{3}}} I(S_{m_2} > t_1). $$
\item We estimate the twin recurrence risk $$\frac{ S(t_{m_2}^{X=0} > t_1 | t_{m_1}^{X=0} > t_1) }{  S(t_{m_2}^{X=0} > t_1) }. $$ The standard errors are e.g. found using Wald estimators for relative risks.
\end{itemize}
\subsubsection*{Step 3. Simulate survival until time $t_1$.}
\begin{itemize}
\item To make the simulations realistic, let the life-time risk of disease be selected from a new sample. 
\item Select $n_4$ unexposed subjects. 
\item For each subject $i$ in $1,2,...,n_4$, we randomly draw the  $U_i \sim \text{Gamma}(1,\nu)$.
\item Similar to Step 1 we find $S_i$ for each $i$. Then we simply estimate $Pr(S_i > t_1)$ with standard confidence intervals, assuming a normal distribution. 
\end{itemize}
\subsubsection*{Step 4. Simulate survival until time $t_1$.}
\begin{itemize}
\item Based on the confidence intervals derived from Part 1-3, we can find the confidence interval of the causal hazard ratio numerically. To do this, we use the strategy suggested in Appendix \ref{appendix: confidence intervals}.
\end{itemize}

\subsection{Validation of particular scenarios}
The simulation algorithm can be used to check the approach under particular scenarios. To illustrate, let the effect of the exposure $X$ conditional on $U$ be a hazard ratio $r_{cau}(t) = 0.80$. Furthermore, let the distribution of $U$ be characterized by mean$=1$ and $\nu = \frac{1}{9}$. The basic hazard rate is set to $h_0(t)= 0.002$. Using the derivations in Section \ref{sec: using family}, we find that the exact value $TRR(t_1) = 1.029$ and $S(t_1^{X=0}) = 0.931)$. This scenario was simulated 500 times. In each simulation, $10^4$ unexposed subjects and $10^4$ exposed subjects were included. Follow-up was initiated at $t_1 = 50$, i.e.\ only those who survived until the age of 50 were included. The duration of follow up was 1 year. This allowed for the calculation of marginal HR from a Cox proportional hazards model. The marginal estimates ($\hat{r}_{mar}(t_1)$) were biased towards the null; Only 31.2\% of the marginal 95\% confidence intervals covered $r_{cau}$. To obtain adjusted estimates, we used simulated data from an independent sample of $10^4$ twin pairs, in which the $TRR(50)$ was estimated. We also used independent sample of $10^4$ subjects to estimate $S(50)$ to obtain frailty adjusted estimates, and confidence intervals were found by the approach in Appendix \ref{appendix: confidence intervals}. These adjusted estimates were better calibrated; 94.8 \% of the adjusted 95\% confidence intervals covered $r_{cau}$. A summary of the simulations is found in Table \ref{table: simulation table} (Scenario 1). 

Similarly, two more extreme scenarios were also simulated (Table \ref{table: simulation table}, Scenario 2 and Scenario 3).  In these scenarios, the same number of subjects were included for estimating $\hat{r}_{mar}(t_1)$, $\text{TRR}(t_1)$ and $S(t_1)$ as in Scenario 1. However, different values for $r_{cau}$, $h_0(t)$ and $\nu$ were introduced. In these scenarios, the coverage of the confidence intervals for $\hat{r}_{mar}(t_1)$ were close to 0, but the adjusted estimates showed coverage approximately $95\%$.  

In all scenarios, the adjusted estimates ($\hat{r}_{adjusted}(t_1)$ in Table \ref{table: simulation table}) are closer to the true value.

\begin{table}[ht]
\centering
\begin{tabular}{lrrrr}
& Scenario & 1 & 2 & 3 \\ 
  \hline
\multirow{5}{*}{\textit{Input}} & Simulations & 500 & 500 & 500 \\ 
 & $r_{cau}$ & 0.80 & 0.70 & 0.70  \\ 
  & $h_0$ & 0.002 & 0.003 & 0.030 \\ 
  & $\nu$ & $\frac{1}{9}$ & $\frac{1}{15}$ & $\frac{1}{5}$ \\ 
&  $t_1$ & 50 & 50 & 50 \\ 
  \hline
\multirow{4}{*}{\textit{Output}} & Median $\hat{r}_{mar}(t_1)$ & 0.89 & 0.89 & 1.01 \\ 
 & Median $\hat{r}_{adjusted}(t_1)$ & 0.81 & 0.70 & 0.71 \\ 
&  Coverage 95\% CI mar & 0.406 & 0.002 & 0 \\ 
&  Coverage 95\% CI adjusted & 0.964 & 0.962 & 0.950 \\ 
\end{tabular}
\label{table: simulation table}
\caption{Summary of 3 simulated scenarios.}
\end{table}


\clearpage

\subsection{Calculation of IV estimates}
\label{sec: IV estimates example}
In many MR analyses, only the association between the instrument and the outcome is reported \citep{burgess2015commentary}. This may be due to the fact that deriving and causally interpreting IV estimates require some caution \citep{vanderweele2014methodological}, in particular for time to event outcomes \citep{burgess2015commentary,tchetgen2015instrumental}. Among other things, the genetic instruments are fixed throughout life, whereas the exposure may be a dynamic process. Indeed, we may interpret the results in Section \ref{sec: standard alcohol example} as an MR analysis, in which only the association between ADH1B (the instrument for alcohol) and all-cause mortality is included. 

\begin{align*}
\hat{\beta}_a &= \frac{\log[\hat{\text{HR}}_{cau}(75)]}{b_{g, [t_1, t_1 + \delta]}(g_1 - g_2)} \nonumber \\
&= \frac{\log(0.52)}{-0.172*14.2(1-0)} \nonumber \\
&= 0.27 \text{ } [0.11, 0.41]
\end{align*}
We may interpret $e^{\hat{\beta}_a}=1.31 \text{ } [1.11, 1.50]$ as the estimated HR under an intervention in which the alcohol consumption was increased by 1 unit per week throughout life. If $\text{HR}_{cau}(75)$ was replaced by $r_{mar}(75)$ in Equation \eqref{expression:identification}, we would yield a HR of $1.17 \text{ } [1.06, 1.29]$. 


This crude example suggests that the magnitude of survival bias may be considerable, in particular when there is an extreme selection of the oldest old. These particular IV results, however, should be interpreted with some caution. I have used summary data from three different sources to obtain the estimates. This illustrates how the procedure can be performed simply by using summary data. However, the results rely on the validity of each source of data. In particular, I have used data from monozygotic twins in Scandinavia to derive the TRR, whereas the study was performed in Australia. The effect of alcohol on mortality may also be non-linear in real life, e.g. that heavy drinking has a larger effect on mortality. Nevertheless, this example illustrates how the frailty approach can be applied to explore bias in IV settings. If specific estimates of $TRR(t_1)$ and $S(t_1)$ are unavailable, the suggested approach can still be performed as a sensitivity analysis: Applied researchers can check the sensitivity to survival bias in their analysis by fitting frailty distributions with various values of $TRR(t_1)$ and $S(t_1)$.

\clearpage
\section{Appendix: R code}
\label{appendix R code}
\begin{lstlisting}[language=R, breaklines=true]
######## Functions ########################
#Find V implements the derivations in Section 2.2 for the Gamma distribution
#Insert FRR (familial recurrence risk or sibling recurrence risk), 
#the cumulative survival S, the observed hazard ratio r_obs is denoted observedHR
findV <- function(FRR,S,observedHR=1.23,from1=1e-30, to1=20,len=30,rr=2.10,plotAndPrint=F,nonStandard=F,printV=F){
  sequence <- seq(from=from1,to=to1,length.out=len)
  rightSide <- FRR*S^2
  tryCatch({
      v <- uniroot(combinedFRRS, S=S,rightSide=rightSide, interval=c(0,to1))$root[1]
    },   warning = function(war) {
      print("warning!!!")
      v <- NA
    }, error = function(err) {
      print("error!!!")
      print(err)
      v <- NA
    }
  ) 
  if(printV){
    print(v)
  }
  A <- v*(1-S^(1/v))/S^(1/v)
  varZ <- 1/v
  causalRiskRatio <- returnCausalRisk(HRobs=observedHR, A=A, v=v)
  if(plotAndPrint==T){
    plot(rightSide-output~sequence, ylim=c(min(0,min(rightSide-output)),max(rightSide-output)), type = "l",xlab="value of v", ylab="S^2*FRR-output")
    abline(a=0,b=0,col=1,lty=2)
    print(paste("The variance value is: ",1/sequence[index],", with difference", rightSide-output[index]))
    print(c(sequence[index],A,riskRatio,causalRiskRatio))
  }
  return(causalRiskRatio)
}

#combinedFRRS returns expression (10)
combinedFRRS <- function(v,S,rightSide){
  denominator <- 1 + 2*(1-S^(1/v))/S^(1/v)
  return ((1/denominator)^v - rightSide)
}

#findVgeneral extends findV to several functions of the PVF family, i.e. the Inverse Gaussian, the Compound Poisson and the Hougaard Family of distributions
findVgeneral <- function(FRR,S,observedHR=1.23,from1=1e-6, to1=100,len=20000,invGaus=F,gamDist=F,comPois=F, Hougaard=F, nonSuscep=0.01,mHougaard=-0.25,rr=2.10,plotAndPrint=F){
  sequence1 <- seq(from=from1,to=to1,length.out=len)
  if(invGaus){
    m = -0.5
    p <- sequence1/m
    sequence <- cbind(sequence1,p,rep(m,length(sequence1))) 
  }
  if(comPois){
    p = -log(nonSuscep)
    m <- sequence1/p
    sequence <- cbind(sequence1, rep(p,length(sequence1)), m)
  }
  if(gamDist){
    p = 1e13; m=1/p
    sequence <- cbind(sequence1,rep(p,length(sequence1)),rep(m,length(sequence1)) )
  }
  if(Hougaard){
    m = mHougaard
    p <- sequence1/m
    sequence <- cbind(sequence1,p,rep(m,length(sequence1))) 
  }
  As <- apply(sequence,1,findAgeneral,S=S) 
  vpmA <- cbind(sequence,As) 
  rightSide <- log(FRR)
  output <- apply(vpmA,1,findDifferenceGeneral,FRR=FRR) # output[1]
  index <- which.min(abs(output-rightSide))
  if(plotAndPrint==T){
    plot(rightSide-output~sequence[,1], ylim=c(min(0,min(rightSide-output)),max(rightSide-output)), type = "l",xlab="value of v", ylab="S^2*FRR-output")
    abline(a=0,b=0,col=1,lty=2)
    print(paste("v: ",vpmA[index,1],"p: ",vpmA[index,2]," m: ",vpmA[index,3]," A: ",vpmA[index,4] ))
    print(sOfT(vpmA[index,]))
  }
  if(comPois || invGaus || Hougaard) {
    v <- vpmA[index,1]; p <-  vpmA[index,2];  m <- vpmA[index,3]
    causalRatio <- as.numeric(returnCausalRisk(HRobs=observedHR, A=As[index], v=v, m=m))
    return(causalRatio[which(causalRatio > 0)])
  }
  else return(paste("The v value is: ",vpmA[index,1],", with difference", rightSide-output[index]))
}

findInverseGaussianRatio <- function(FRR,S, to1=20,len=1000,rr=2.10,observedHR=1.23){
  m <- -0.5
  rightHand <- 2
  sequence <- cbind(sequence1,p,rep(m,length(sequence1))) 
}
#findInverseGaussianV <- function(v,A)


#findAgeneral H_0 (which is denoted A in the function), given data on \nu, \rho and m
findAgeneral <-function(S,vpm){
  v <- vpm[1]; p = vpm[2]; m = vpm[3]
  u <- (log(S)/p +1)^(1/m)
  A <- (v-u*v)/u
  return(A)
}

#findDifferenceGeneral is used to find the minimal \nu in findVgeneral
findDifferenceGeneral <- function(inputVector,FRR){
  v <- as.vector(inputVector)[1]; p <- as.vector(inputVector)[2]; m <- as.vector(inputVector)[3]; A <- as.vector(inputVector)[4]
  return(p*(1-2*(v/(v+A))^m + (v/(v+2*A))^m ))
}

#riskRatio6.15 is derived from formula 6.15 in Aalen et al, Survival and Event history analysis
riskRatio6.15<- function(p,m,v,A,r){
  nominator <- r * (1 + A / v)^(m + 1)
  denominator <- (1 + r*A / v)^(m + 1)
  return(nominator/denominator)
}

#returnCausalRisk is used to numerically find the causal hazard ratio, given r_obs (HRobs)
returnCausalRisk <- function(HRobs, A, v,m=0){
  r=NA
  if(m==0){
    r <- HRobs / (1 + A/v - HRobs*A/v )
  } else if(m==-0.5){
    r <- polyroot(c(-HRobs^2,-HRobs^2*A/v,1+A/v))
  } else r <- tryCatch({
    uniroot(HRcausalFunction, interval=c(0,6), HRobs=HRobs, A=A, v=v, m=m)[1]
  },   warning = function(war) {
    print("warning!!!")
    return(0)
  }, error = function(err) {
    print("error!!!")
    return(0)
  }
  ) 
  #r <- HRobs^(1/(m+1)) / (1 + A/v - HRobs^(1/(m+1))*A/v )
  return(r)
}

#returnCausalRisk is used to numerically find the causal hazard ratio, given r_obs (HRobs)
#This is expression (13)
returnObservedRisk <- function(r, A, v,m=0){
 HRobs <- r * ( (1 + A/v) / (1 + r*A/v) ) ^ (m + 1)
return(HRobs)
}


#HRcausalFunction is used in findVgeneral to numerically find the causal hazard ratio, given r_obs (HRobs)
HRcausalFunction <- function(r, HRobs, A, v, m){
  equationHR <- HRobs^(1/(m+1))*(1 + A*r/v) - r^(1/(m+1))*(1+A/v)
  return(equationHR)
}

#Finding the FRR ov Y=1, given data on Y=0 (for figure 2)
frrEventFunction <- function(S, FRR){
  (FRR*(1-2*S+S^2)-1+2*S)/S^2
}

# Veryfying that the code is OK
sOfT <- function(vpmA){
  v = vpmA[1]; p = vpmA[2]; m = vpmA[3]; A = vpmA[4]
  withinLog <- -p*(1-(v/(v + A))^m)
  return(exp(withinLog))
}

# We can now derive numeric confidence intervals by simulations. 
# Let k denote the number of confidence intervals. 
numericCIgamma <- function(HRobsDerived=resultsNaiveCoxph,RiskDerived=estimatedPropGeneral,FRRderived=riskRatioInterval){
  iLogHRobs <- rnorm(1,mean = log(HRobsDerived[1]), sd = (log(HRobsDerived[1]) - log(HRobsDerived[2]))/qnorm(0.975))
  iS <- rnorm(1,mean = RiskDerived[1], sd= abs(RiskDerived[1]-RiskDerived[2])/qnorm(0.975))
  iLogFRR <- rnorm(1,mean = log(FRRderived[1]), sd= (log(FRRderived[1])-log(FRRderived[2]))/qnorm(0.975))
  output <- findV(S=iS,FRR=exp(iLogFRR),observedHR=exp(iLogHRobs),nonStandard = T,printV=F) 
  return(output)
}

numericCIpvf <- function(HRobsDerived=resultsNaiveCoxph,RiskDerived=estimatedPropGeneral,FRRderived=riskRatioInterval,invGausIn=F,gamDistIn=F,comPoisIn=F, HougaardIn=F,nonSuscepIn=0.01,mHougaardIn=-0.25,rrIn=2.10){
  iLogHRobs <- rnorm(1,mean = log(HRobsDerived[1]), sd = (log(HRobsDerived[1]) - log(HRobsDerived[2]))/qnorm(0.975))
  iS <- rnorm(1,mean = RiskDerived[1], sd= abs(RiskDerived[1]-RiskDerived[2])/qnorm(0.975))
  iLogFRR <- rnorm(1,mean = log(FRRderived[1]), sd= (log(FRRderived[1])-log(FRRderived[2]))/qnorm(0.975))
  output <- findVgeneral(S=iS,FRR=exp(iLogFRR),observedHR=exp(iLogHRobs),invGaus=invGausIn,gamDist=gamDistIn,comPois=comPoisIn, Hougaard=HougaardIn, nonSuscep=nonSuscepIn,mHougaard=mHougaardIn,rr=rrIn) 
  return(output)
}


######## Functions finished #################


######## Numeric calculations follows #######
# Derivation of confidence intervals:
# Since r is a monotone function of r_obs (empirically justified by plots), 
# we plug in the bounds of r_obs and solve expression (10)

#robs = 1.5; surviving = 0.9 st
robs = 1.2; surviving = 0.5
###### Explicit calculations ################
#Figure 2
FRRs <- seq(from=1.11, to=5, length.out=501)
FRRs1s <- sapply(FRRs,frrEventFunction,S=surviving)
FRRs1s <- seq(from=1.03, to=1.4, length.out=501)
estim <- sapply(FRRs1s,findV,S=surviving,observedHR=robs)  
estim2 <- sapply(FRRs1s,findVgeneral,S=surviving,invGaus = T,observedHR=robs) 
estim3 <- sapply(FRRs1s,findVgeneral,S=surviving,Hougaard = T,mHougaard = -0.125,observedHR=robs) 
estim4 <- sapply(FRRs1s,findVgeneral,S=surviving,comPois=T,nonSuscep = 0.01,observedHR=robs)
#plot(estim~FRRs,type="l",ylab="Hazard ratio",xlab="Familial Relative Risk (FRR)",ylim=c(1,5),xlim=c(1,5))
#abline(a=1.5,b=0,lty=2)

plot(estim~FRRs1s,type="l",ylab="Hazard ratio",xlab="Twin Relative Risk (TRR)",ylim=c(1,3),xlim=c(1.1,1.4))
abline(a=1.2,b=0,lty=2)

lines(FRRs1s,estim2,col=2)
lines(FRRs1s,estim3,col=3)
lines(FRRs1s,estim4,col=4)
legend(x="topleft",c(expression('HR'[cau]*'(90), compound Poisson'),expression('HR'[cau]*'(90), Gamma'),expression('HR'[cau]*'(90), Hougaard'),expression('HR'[cau]*'(90), inverse Gaussian'), expression('HR'[MR]*'(90)')),lty=c(1,1,1,1,2), col=c(4,1,3,2,1),cex=1.25)


### New Section - see how bias varies with follow up time. 
#Let the rate be constant alpha
propAlive <- seq(from=1, to=0, length.out = 1001)
#Assume distriubtions with var=1
rIn=1.2
gammaVpm <- c(1,1e9,1/1e9); invGausVpm <- c(0.5,-1,-0.5); hougaardVpm <- c(-0.125+1,(-0.125+1)/-0.125 ,-0.125); compPoissonVpm <-c(1/(-log(0.1)-1)+1, -log(0.1), 1/(-log(0.1)-1) )
cumHazardsGamma <- findAgeneral(propAlive,gammaVpm)
cumHazardsInvGaus <- findAgeneral(propAlive,invGausVpm)
cumHazardsHougaard <- findAgeneral(propAlive,hougaardVpm)
cumHazardsCompPois <- findAgeneral(propAlive,compPoissonVpm)


observedRisksGamma <- sapply(cumHazardsGamma, returnObservedRisk, r=rIn,v=gammaVpm[1], m=gammaVpm[3] )
observedInvGaus <- sapply(cumHazardsInvGaus, returnObservedRisk, r=rIn,v=invGausVpm[1], m=invGausVpm[3] )
observedHougaard <- sapply(cumHazardsHougaard,returnObservedRisk, r=rIn,hougaardVpm[1], m=hougaardVpm[3] )
observedCompPois <- sapply(cumHazardsCompPois,returnObservedRisk, r=rIn,compPoissonVpm[1], m=compPoissonVpm[3] )

plot(observedRisksGamma~propAlive, xlim=c(1,0),ylim=c(0.7,1.3),type="l", xlab="Proportion alive", ylab = "Unadjusted hazard ratio", lty=2)
lines(observedInvGaus ~ propAlive,lty=2, col=2)
lines(observedHougaard ~ propAlive,lty=2, col=3)
lines(observedCompPois ~ propAlive,lty=2, col=4)
abline(a=rIn,b=0,lty=1)
legend(x="bottomleft",c(expression('HR'[obs]*'(90), compound Poisson'),expression('HR'[obs]*'(90), Gamma'),expression('HR'[obs]*'(90), Hougaard'),expression('HR'[obs]*'(90), inverse Gaussian'), expression('HR'[cau]*'(90)')),lty=c(2,2,2,2,1), col=c(4,1,3,2,1),cex=1.25)

returnObservedRisk(r, A, v,m=0)

#Example on Alcohol and all cause mortality, Section 4
frrAussie <- 1.27
sAussie <- 0.56
obsAussie <- 0.68; lowerAussie <- 0.54; upperAussie <- 0.87

#Results in Section 4.1 
aussieEstimates <- c(obsAussie,lowerAussie,upperAussie) #aussieEstimates <- c(3.0,2.0,4.0)
findVgeneral(FRR = frrAussie, S=sAussie, observedHR = lowerAussie, invGaus = T)
aussieGamma <- sapply(aussieEstimates,findV,FRR = frrAussie, S=sAussie,nonStandard = T,printV=T)
aussieInvGaus <- sapply(aussieEstimates, findVgeneral,S=sAussie,FRR=frrAussie,invGaus = T, plotAndPrint = T,to1=1)
aussieHougaard <- sapply(aussieEstimates, findVgeneral,S=sAussie,FRR=frrAussie,Hougaard = T,mHougaard = -0.125, plotAndPrint=T,to1=1)
aussieCPound <- sapply(aussieEstimates, findVgeneral,S=sAussie,FRR=frrAussie,comPois=T,nonSuscep = 0.10,plotAndPrint=T)

#Find numeric confidence intervals.
set.seed(1)
CIalmeidaExample <- replicate(n=2e3,numericCIgamma(HRobsDerived=c(0.68,0.54,0.87),RiskDerived=c(0.56,0.56,0.56),FRRderived=c(1.27,1.20,1.34)))
quantile(CIalmeidaExample,c(0.025,0.5,0.975))
CIalmeidaExampleInvGau <- replicate(n=2e3,numericCIpvf(HRobsDerived=c(0.68,0.54,0.87),RiskDerived=c(0.56,0.56,0.56),FRRderived=c(1.27,1.20,1.34),invGausIn=T))
quantile(CIalmeidaExampleInvGau,c(0.025,0.5,0.975))
CIalmeidaExampleComPois <- replicate(n=2e3,numericCIpvf(HRobsDerived=c(0.68,0.54,0.87),RiskDerived=c(0.56,0.56,0.56),FRRderived=c(1.27,1.20,1.34),comPoisIn =T,nonSuscepIn = 0.10 ))
quantile(CIalmeidaExampleComPois,c(0.025,0.5,0.975))
CIalmeidaExampleHougaard <- replicate(n=2e3,numericCIpvf(HRobsDerived=c(0.68,0.54,0.87),RiskDerived=c(0.56,0.56,0.56),FRRderived=c(1.27,1.20,1.34),HougaardIn = T,m = -0.125 ))
quantile(CIalmeidaExampleHougaard,c(0.025,0.5,0.975))


#Results in Section 4.1.1
exp(log(0.68) / (-0.172*14.2)) #Per unit per week. Observed


######## Calculations finished ##################
######## The next section describes simulations #
######## New functions are introduced, and ######
######## and the former functions will also #####
######## be used. ###############################

library(survival)
library(epitools)
# We will now show how simulations can be used to verify the approach. 
# n number of individuals, h0 is baseline hazard
# nu is parameter of gamma, t1 time at follow up
# Assume that we observe individuals until time t1=50 years. Who survive?
# Let the HR in the exposed individuals be 0.8
# Among the survivors, we can now estimate the survivor times. Let H(t) = ht. 
# The inverse is H-1(H(t)) = H(t)/h = t. I.e. an Exponential distribution.

# We will now simulate the hazard function

sampleSurvivalTime <- function(u,h0,r=1){
  uni <- runif(1)
  out <- - log(uni) / (h0*u*r)
  return(out)
}

# Let us assume a Gamma distributed frailty
# Let the rate=nu such that t1 is the time to start follow up
# and tFollow is the number of years under follow up time
getConfidenceIntervalGamma <- function(n=1e6, h0=0.002, nu=1/9,HRcau=0.8,t1=50,tFollow = 2){
  uExp <- rgamma(n, shape=nu, scale = 1/nu )
  uUnexp <- rgamma(n, shape=nu, scale = 1/nu )
  
  #Sample survival times
  timesExp <- sapply(uExp, sampleSurvivalTime, h0=h0, r=HRcau) #summary(timesExp)
  timesUnexp <- sapply(uUnexp, sampleSurvivalTime, h0=h0, r=1) #summary(timesUnexp)
  
  # Select those individuals who have survival times larger than t1
  survivorsTimesExp <- timesExp[timesExp > t1] - t1
  survivorsTimesUnexp <- timesUnexp[timesUnexp > t1] - t1
  eventsExp <- survivorsTimesExp < tFollow
  eventsUnexp <- survivorsTimesUnexp < tFollow
  modifsurvivorsTimesExp <- survivorsTimesExp; modifsurvivorsTimesExp[ modifsurvivorsTimesExp > tFollow] <- tFollow
  modifsurvivorsTimesUnexp <- survivorsTimesUnexp; modifsurvivorsTimesUnexp[ modifsurvivorsTimesUnexp > tFollow] <- tFollow
  
  # Make input ready for the coxph model
  exposureIndicator <- c(rep(1,length(modifsurvivorsTimesExp)), rep(0,length(modifsurvivorsTimesUnexp)))
  statuses <- c(eventsExp,eventsUnexp)
  times <- c(modifsurvivorsTimesExp, modifsurvivorsTimesUnexp)
  
  # Let us fit a Cox model 
  naiveCoxph <- coxph(Surv(times, statuses) ~ exposureIndicator)
  resultsNaiveCoxph <- (summary(naiveCoxph)$conf.int)[c(1,3,4)]
  
  # First, assume that the survival is known. Then we simply use Expression 5 to find S
  St1 <- (nu/(nu+h0*t1))^nu
  # Furthermore, we use expression 8 to find the FRR(t1)
  FRRt1 <- (nu/(nu+2*h0*t1))^nu / St1^2
  # Then we use the functions to find the observed value
  adjustedCoxph <- findV(S=St1,FRR=FRRt1,observedHR=resultsNaiveCoxph,nonStandard = T,printV=F) 
  #This looks very good....!
  return(rbind(resultsNaiveCoxph,adjustedCoxph))
}

# Now, assume that S and FRR are only known with uncertainty
# Assume that we obtained n1 unexposed twin pairs from the same population with Us. 
findExpectedSurvival <- function(h,t,u,hr=1){
  H <- h*t
  return(exp(-H*u*hr))
}

# Function to find the expected survival until time t1
getCIgeneralPopulation<- function(n1In=n1, nuIn=nu,h0In=h0,tIn=t1,hrIn=HRcau){
  uGeneralPopulation <- rgamma(n1In, shape=nuIn, scale = 1/nuIn)
  probSurvivalGP <- findExpectedSurvival(h0In,t=tIn,hr=1,u=uGeneralPopulation) # Rather than hr=hrIn
  survivorsGP <- sapply(probSurvivalGP, rbinom,n=1,size=1)
  propGP <- prop.test(table(survivorsGP))
  estimatedPropGP <- 1-c(propGP$estimate, propGP$conf.int)[c(1,3,2)]
  return(estimatedPropGP)
}

# Function to find the twin recurrence risk
getCItwinRecurrenceRisk <- function(n1In=n1, nuIn=nu,h0In=h0,tIn=t1,hrIn=HRcau){
  uTwinPairs <- rgamma(n1In, shape=nuIn, scale = 1/nuIn)
  probSurvival <- findExpectedSurvival(h0In,t=tIn,hr=1,u=uTwinPairs) # rather than hr=hrIn
  survivorsTwinPairs <- cbind(sapply(probSurvival, rbinom,n=1,size=1),sapply(probSurvival, rbinom,n=1,size=1))
  survivorsConditional <- survivorsTwinPairs[survivorsTwinPairs[,1]==1,2]
  # For a straightforward approach, let us use the first column for the estimation of the Survival. 
  # Find the proportion of successes
  generalProb <- prop.test(table(survivorsTwinPairs[,1]))
  estimatedPropGeneral <- 1-c(generalProb$estimate, generalProb$conf.int)[c(1,3,2)]
  conditionalProb <- prop.test(table(survivorsConditional))
  estimatedProbConditional <- 1-c(conditionalProb$estimate, conditionalProb$conf.int)[c(1,3,2)]
  tableMatrix <- as.table(rbind(table(survivorsTwinPairs[,1]),table(survivorsConditional)))
  riskRatioInterval <- riskratio.wald(tableMatrix)$measure[2,]
  return(riskRatioInterval)
}

# To check if the confidence interval is correct, we repeat calculations of the observed HR, the FRR and the Survival probability
checkCIisCorrect <- function(nInput=1e4,nInputTwin=1e4, h0Input=0.002, nuInput=1/9,HRcauInput=0.8,t1Input=50,tFollowInput = 1, k=1e4){
  CIHRobserved <- getConfidenceIntervalGamma(n=nInput, h0=h0Input, nu=nuInput,HRcau=HRcauInput,t1=t1Input,tFollow = tFollowInput)
  CIriskGP <- getCIgeneralPopulation(n1In=nInput, nuIn=nuInput,h0In=h0Input,tIn=t1Input,hrIn=HRcauInput)  
  CIriskTwins <- getCItwinRecurrenceRisk(n1In=nInputTwin, nuIn=nuInput,h0In=h0Input,tIn=t1Input,hrIn=HRcauInput)
  #print(CIHRobserved[1,])
  #print(CIriskGP)
  #print(CIriskTwins)
  totalDraw <- replicate(n=k,numericCIgamma(HRobsDerived=CIHRobserved[1,],RiskDerived=CIriskGP,FRRderived=CIriskTwins))
  CItotal <- round(quantile(totalDraw, c(0.025,0.5, 0.975)),2)
  return(c(CItotal,CIHRobserved[1,]))
}


######## Explicit input to simulations ####
set.seed(123)
output1st123 <- replicate(n=5e2,checkCIisCorrect())
# Check how many CI that cover the true value: 
#output1st123 <- get(load("/Users/matsjuliusstensrud/Documents/NEJM/data/simDataEpiMethods1version2.RData"))
str(output1st123)
sum(output1st123[1,] < 0.8 & output1st123[3,] > 0.8)
Toutput1st123 <- t(output1st123)
coverage1stAdjusted <- 1-sum(Toutput1st123[,1] > 0.80 | Toutput1st123[,3] < 0.80) / 500 # Of these, 0.092 coverage. 0.91
coverage1stMar <- 1-sum(Toutput1st123[,5] > 0.80 | Toutput1st123[,6] < 0.80) / 500 # Of these, all of the intervals are above 
summary(Toutput1st123[,2])
summary(Toutput1st123[,4])


set.seed(123)
output2nd123 <- replicate(n=5e2,checkCIisCorrect(nInput=1e6, h0Input=0.003, nuInput=1/15,HRcauInput=0.7,t1Input=50,tFollowInput = 1, k=1e5))
sum(output2nd123[1,] < 0.7 & output2nd123[3,] > 0.7)
summary(output2nd123[2,])
Toutput2nd123 <- t(output2nd123)
coverage2ndAdjusted <- 1-sum(Toutput2nd123[,1] > 0.70 | Toutput2nd123[,3] < 0.70) / 500. # Of these, 0.092 coverage. 0.91
coverage2ndMar <- 1-sum(Toutput2nd123[,5] > 0.70 | Toutput2nd123[,6] < 0.70) / 500. # Of these, all of the intervals are above 
summary(Toutput2nd123[,2])
summary(Toutput2nd123[,4])



set.seed(123)
output3rd123 <- replicate(n=5e2,checkCIisCorrect(nInput=1e6, h0Input=0.030, nuInput=1/5,HRcauInput=0.7,t1Input=50,tFollowInput = 1, k=1e5))
Toutput3rd123 <- t(output3rd123)
coverage3rdAdjusted <- 1-sum(Toutput3rd123[,1] > 0.70 | Toutput3rd123[,3] < 0.70) / 500. # Of these, 0.092 coverage. 0.91
coverage3rddMar <- 1-sum(Toutput3rd123[,5] > 0.70 | Toutput3rd123[,6] < 0.70) / 500. # Of these, all of the intervals are above 
summary(Toutput3rd123[,2])
summary(Toutput3rd123[,4])

\end{lstlisting}

\clearpage

\clearpage

\clearpage

\clearpage
\end{document}